\documentstyle[12pt]{article}
\begin {document}
\newcommand {\beqa}{\begin{eqnarray}}
\newcommand {\eeqa}{\end{eqnarray}}
\newcommand {\n}{\nonumber \\}
\newcommand {\beq}{\begin{equation}}
\newcommand {\eeq}{\end{equation}}
\newcommand {\om}{\omega}
\newcommand {\s}{\sigma}

\newcommand {\pa}{\partial}

\newcommand {\th}{\theta}

\begin{flushright}
RI-7
\\
June  1994
\end{flushright}
\begin{center}
{\large\bf Integrable Vector Perturbations of W-invariant Theories \\
              and \\
      their Quantum Group Symmetry} \vspace{.7 in}\\
{\bf A. Babichenko
        \vspace{.4 in}\\
        {\it Racah institute of Physics, Hebrew University }\\
        {\it Jerusalem, 91904, Israel  \vspace{.7 in}}}
\end{center}
\begin{abstract}
  Perturbations of $WD_n$ and $W_3$ conformal theories which
 generalize the
$(1,2)$ perturbations of conformal minimal models are shown to be
integrable by counting argument.  $A_{2n-1,q}^{(2)}$ and $D_{4,q}^
{(3)}$ symmetries of corresponding S-matrices are conjectured and
proved by explicit construction of conserved nonlocal charges in the
$WD_3$ case with the proper quantum group of symmetry.
\end{abstract}
\newpage

{}~
\section{Introduction}

     In the recent years a progress has been achieved in the
understanding of integrable perturbations of two dimensional
 Conformal Field Theories
(CFT) from the point of view of their current algebra and Factorizable
Scattering Theory (FST) of perturbed massive model. It was realized
from
the very beginning [1], that the structure of the (1,2)
perturbation
of minimal CFT,corresponding on the classical level to
Zhiber-Mikhailov-Shabat model, and its FST
is more complicated then  the structure of of (1,3) perturbation,which
classically
corresponds to Sine-Gordon model.  After the work of
Fateev and Zamolodchikov [2] it was not clear whether there
exists some
general description of FST and a symmetry of (1,2) perturbations of
minimal models, because the discovered group of symmetry of FST
turned out
to be  very different: $E_8$,$E_7$ and $E_6$ for $p$=3,4 and 6
minimal models
correspondingly. Nevertheless  the  general solution with
$A_{2q}^(2)$ symmetry
of FST of any (1,2) perturbed  minimal model was found by Smirnov
[3]
which reproduces the previously known solutions as particular cases,
 however
this reproduction is essentially nontrivial and is based on
the properties of
representations of $ sl(2)_{q} $ at special values of q equal to
root of unity.

    Along with integrable perturbations of conformal models
with Virasoro and
 Kac-Moody algebra, perturbations of CFT with other additional
symmetries and their
FST were studied. In ref. [4]
 few integrable perturbations of CFT $Z_N$ parafermionic
models were studied, which are the lowest minimal models of
$W$-invariant
theories.  The (1,3) integrable perturbation of CFT was naturally
generalized
to $W$-invariant theories as a perturbation by the field
corresponding to the
 adjoint representation of the algebra $A_n$, and their FST were
constructed in [5]. In [4] and [5] the
vacuum structure
of the theory was conjectured to be in  correspondence
 to the admissibility diagram of some
Interaction Round the Face (IRF) models, and in [5]
the S-matrix of
the model was
 explicitly constructed by  solving of Yang Baxter equation
 by use of
 $A_n$ invariant IRF Boltzmann weights. The integrability of
such a perturbation of an adjoint type for $WX$-invariant
theories constructed
 on an arbitrary Lie group $X$ ( treated as the coset construction
$\frac{X_p\times X_1}{X_{p+1}}$ ,see, for example, [6] )
is known for a long time, and some examples of
corresponding FST for these models ($B_n,C_n,D_n$ and few others)
were discussed, for example, in few recent works of
Gepner [7]. On the classical level the trivial reason
for integrability
of the adjoint type perturbation is expressed by the fact
of correspondence
of $\phi_{adj}$ to the maximal positive root of the algebra such
that together
with screening operators of the $WX$-invariant theory
 they form $X$-invariant Affine Toda Field Theory (ATFT) with
imaginary coupling constant [8]. In this sense the
(1,2) perturbation of Virasoro minimal models may be called
vector type
perturbation (with respect to $sl(2)$).

    The natural question arrives: are there generalizations
of (1,2) type
integrable perturbations of CFT to W-invariant theories?
Recently the existence
of such perturbations was
pointed out [9]. It was conjectured that,
 $WD_n^{(p)} + \phi_{vect}$ for
$n\geq 3$ is integrable, where $\phi_{vect}$ is the primary
field corresponding
to the fundamental weight of vector representation of $D_n$ (the field
$(211...1|11...1)$ in the notation of ref.[6] ). The hint
 for the
integrability of such perturbed model is actually seen
even on the "classical"
level, since the perturbing field together with screening operators
of the $WD_n$ give rise to the $B_n$ imaginary coupled ATFT.
 It was conjectured
that the FST of this integrable model should have
 $A_{2n-1,q}^{(2)}$ symmetry.
For $n=2$ it was found that the corresponding analog of
this integrable vector
perturbation is $WA_{2}^{(p)} + (21|11)$, and $D_{4}^{(3)}$
symmetry of
corresponding
 FST was conjectured. In this case the Hamiltonian of perturbed model
completed by screening operators of $WA_2$ (or $W_3$ in usual
notations) form
the $G_2$ imaginary coupled ATFT.

Checks of integrability of $W_3$ and $WD_3$ which
 were done
in [9],  are exact for irrational
values of central charge, since
 using the counting argument
 they did not take into account the summation over
the root lattice
in the formula of W characters, ignoring the highest null vectors.
The suggestion of the  symmetry of the FST theory in this work
 was done analyzing the $p\rightarrow\infty$ and is in
complete correspondence
 with the non simply-laced duality  of real coupled ATFT
observed recently in
refs [10],[11].

 In this work we study in more details the above described
vector perturbations of $WD_n$ (and $W_3$) theories trying
to show explicitly
 the presence of $A_{2n-1,q}^{(2)}$ symmetry with the help
 of nonlocal
currents of the model, and then discuss our attempt to obtain
 the S-matrix on the
base of the $A_{2n-1,q}^{(2)}$-symmetric R-matrix built in the work
[12]. In section 2 we start from more rigorous check of
 integrability
 by counting argument with exact calculation of $W$-characters
at rational
values of central charge
(minimal models). After that (section 3) nonlocal charges for n=3
case  with the algebra $A_{5,q}^{(2)}$ are constructed
explicitly and the
role of the gradation chosen for the R-matrix (as a starting point
for the S-matrix construction), which commute
 with obtained
comultiplication, is discussed. In the last section we summarize the
 results and briefly discuss one other integrable
perturbation of $WD_n$ model.

{}~
\section{Vector perturbations and their integrability}

    Before starting with the hamiltonian of the
vector perturbation for
$W_3$ minimal models in the free field representation,
recall the standard
notations of primary fields (ref [6]) for $WX_n^{(p)}$
 minimal model.
Primary field $\Phi_{(\vec{l}|\vec{l'})}$ is characterized
by the set of
integers $(l_1,...,l_r|l'_1,...,l'_r)$, where $r$ is the rank of $X_n$.
 It can be written as

\beqa
    \Phi_{(\vec{l}|\vec{l'})}= :e^{i\vec{\beta}\vec{\phi(z)}}:
    \label{PrFi}\\
    \vec{\beta}=\sum_{i=1}^{r} (\alpha_{+}(1-l_i)+
    \alpha_{-}(1-l'_i)) \vec{\om}_i \nonumber
\eeqa

where $\alpha_{+}=\sqrt{\frac{p}{p+1}}$,
 $\alpha_{-}=-\sqrt{\frac{p+1}{p}}$,
and $\vec{\om}_i$ are fundamental weights of the algebra $X_n$.
 Screening
fields are  $:e^{i \alpha_{\pm} \vec{\alpha}_{i} \vec{\phi(z)}}$, where
$\vec{\alpha}_i$, $i=1,...,r$, are positive roots of $X_n$.

    Consider Hamiltonian for the perturbation of $W_3$
conformal theory by the operator $\Phi_{(21|11)}$ which
 has the dimension
$\frac{1}{3}(1-\frac{4}{p+1})$ in the $(p,p+1)$ unitary minimal model.
 This Hamiltonian may be written as

\beq
     H=\lambda \int d^{2} z
:(e^{i \alpha_{+}\vec{\alpha}_{1}\vec{\phi(z)}}
 + e^{i \alpha_{+}\vec{\alpha}_{2}\vec{\phi(z)}}
 + e^{-i \alpha_{+}\vec{\om}_{1}\vec{\phi(z)}}):
\label{HamW3}
\eeq

    The set of vectors $(\vec{\alpha}_{1},\vec{\alpha}_{2},
-\vec{\om}_{1})$ expressed, for example, in standard ortonormal
 basis $(e_1-e_2, e_2-e_3, 2/3 e_1 -1/3 e_2 -1/3 e_3)$, obviously
 forms the set of roots of $G_2$ affine algebra. This fact allows
us to
 consider the above perturbed CFT as a good candidate to integrable
 model,which is the $G_2$ ATFT with the imaginary coupling
constant, and its conserved currents should have spins equal to the
exponents of $G_2$ (3,5) modulo its Coxeter number (6).
 But we should be convinced that the integrability survives
 also on the quantum level. The simplest way to try to check quantum
integrability is to check the working of so called counting argument
[1],
which is sufficient but not necessary condition of integrability. It
says that, comparing
dimensions of the representation of the $W$ algebra based on the
 perturbing operator as primary field, modulo derivatives,
 with those of
unity operator on different Virasoro levels, we can see the existence
 of a conserved current of higher spin , and hence the integrability,
 if it turns out that the dimension  of the
former is less then
 the dimension of the latter. The dimensions of Verma moduli
can be extracted from the characters of highest weight representations
of corresponding $W$-algebra. Using the character formula (see, for
 example, ref [6]) for the "completely degenerate"
representations of $W$-algebra

\beqa
    \chi(\vec{\Omega},\vec{\Omega}')&=&\left[ q^(1/24)
\prod_{i=1}^{\infty}(1-q^i)\right] ^{-r}\times \label{char} \\
   &&\times \sum_{\hat{s}\in w} \sum_{\vec{\lambda}\in \Gamma_{\alpha}}
  det(\hat{s}) q^{[p\hat{s}\vec{\Omega}-(p+1)\vec{\Omega}'+p(p+1)
   \vec{\lambda}]^{2}/2p(p+1)} \nonumber
\eeqa

where $(\vec{\Omega},\vec{\Omega}')=(\vec{\om}_i l^{i},
\vec{\om}_i l'^{i})$ is the primary field in the notations of
(\ref{PrFi}),the  sums run over the elements $\hat{s}$ of the Weyl
group $w$ and the root lattice $ \Gamma_{\alpha}$ of the Lie algebra
of the rank $r$, and $p$ is the number of minimal model, we found
(with a  help of Mathematika) that according to their Virasoro levels
$n=2,...10$
the dimensions of Verma moduli of the perturbing field
 $(21|11)$ modulo total
derivatives are $(1,1,2,2,3,4,6,6,10)$ for $p=4$,
 $(1,1,3,3,6,7,13,15,25)$ - for $p=5$, $(1,1,3,3,6,7,13,15,26)$ -for
 $p\geq 6$. The same dimensions calculated for the operator of unity
 ($(11|11)$-field) are $(1,1,1,1,3,1,4,4,6)$ -for $p=4$, and
$(1,1,1,1,4,2,7,7,12)$- for$p\geq 5$. We see, comparing the
 dimensions
 of levels which correspond to spin equal to $5$, that there is a
conserved current of that spin, as it should be according to our
observation on the $G_2$ ATFT structure of the perturbed theory
( 5 is one of the exponents of $G_2^{(1)}$ ). The explicit form of the
conserved charge

\begin{equation}
P_6(z,\bar z)= a:T^3: + b:(\partial T)^2: + c:W^2: + d:W\partial T:
\label{tok}
\end{equation}

was argued in [9], where $T$ and $W$ are energy momentum tensor
and generator of $W$-symmetry, and $a,b,c,d$ are some constants.

    In the same way if we perturb $WD_n^(p)$ minimal  theory by the
 operator $\Phi_{(21...1|11...1)}$ with  the conformal dimension

\beqa
\Delta&=&\frac{1+(np-(n-1)(p+1))^{2}+\sum_{k=2}^{n-2} k^2}{2p(p+1)}
{}~~~~(n\geq 4) \label{Dime} \\
\Delta&=&\frac{1+(p-2)^{2}}{2p(p+1)}
{}~~~~(n= 3) \nonumber
\eeqa

 it easily can be seen, that the set
of  screening vertex operators together with the perturbing one forms
the potential of imaginary coupled $B_n$ ATFT.
The check of counting argument in this
case by use of the formula (\ref{char}) gives the following sequences
 of dimensions
of Verma moduli at different Virasoro levels (spins) $2,...,11$:
 1) $n=3$  $(1,1,2,1,5,4,11,11,22,26)$ -for the unity operator and
 $(2,1,5,5,12,14,28,36,64,85)$ -for the perturbing one;
 2) $n=4$  $(1,0,3,0,5,2,11,7,22,19)$ -for the unity operator and
 $(1,2,3,5,9,13,22,33,52,77)$ for the perturbing operator;
 3) $n=5$  $(1,0,2,1,4,2,9,7,18,18)$ -for the unity and \newline
 $(1,1,4,3,9,10,21,26,48,63)$ -for the perturbing operator.
 So we see the existence of conserved charges of spin 3 for each $n$
 and even  a charge of spin 5 for $n=5$ case, in full correspondence
with the exponents of $B_n$ ATFT.

    As we mentioned in the introduction, the conjectured symmetries of
the factorized S-matrices for these integrable field theories are
 expressed by the algebras which are dual to the corresponding
 non simply
laced ATFT (with an imaginary coupling constant), i.e. $D_{4}^{(3)}$
for $G_2$ and $A_{2n-1}^{(2)}$ -for $B_n$. The presence of these
 symmetries in the perturbed models were argued in [9] on the
basis of the analysis of the conserved (local) charge algebras. In
the next section we shall build explicitly nonlocal charges in vector
perturbed $WD_3^{(p)}$ theory which form the $A_{5,q}^{(2)}$ algebra
and shall discuss its representation by fundamental solitons.

    \section{Algebra of nonlocal charges and its representation}

   Quantum deformed symmetries play the central role in the
investigation of quantum integrable systems  and give a powerful
tool for construction of factorized exact S-matrix for the system
 provided the R-matrix  corresponding to this quantum group symmetry
is known. The way to see some certain quantum symmetry of an integrable
perturbed CFT explicitly in terms of its free field representation
 may lay only in the construction of nonlocal currents expressed in
terms of these free fields, since only braiding relations of the
nonlocal currents may give rise to some quantum deformed algebra.
The construction of nonlocal charges with $ sl(2)_{q} $ algebra of
symmetry
 in Sine-Gordon theory  obtained
as a perturbations of minimal unitary conformal models was done in the
works [13]. There it was shown  how this
construction may be generalized to any ATFT with other group of
symmetry, and in the work [14] the results of Smirnov for the
S-matrix of $(1,2)$ perturbations of minimal models were reproduced
from the point of view of nonlocal charges.
 In this section we shall follow [14] and construct nonlocal
charges for the
integrable model $WD_3^{(p)} + \Phi_{(211|111)}$ which form the algebra
$A_{5,q}^{(2)}$ and show how the the change of gradation for R-matrix
of this group of symmetry (from the homogeneous to the spin gradation
in which we shall observe our quantum symmetry) will fix
the dependence
of the effective coupling constant on the bare coupling $\beta$.

   So, we are going to consider perturbed $WD_3^{(p)}$ theory with the
Hamiltonian

\beqa
 H&=&\frac{\lambda}{2\pi}\int d^{2}z \left( e^{-i\beta(\Phi_1-\Phi_2)}+
   e^{-i\beta(\Phi_2-\Phi_3)} + e^{-i\beta(\Phi_2+\Phi_3)} +
e^{i\beta\Phi_1} \right) \n
&=&\frac{\lambda}{2\pi}\int d^{2}z \Phi_{pert}(z,\bar{z})
\label{HamD}
\eeqa

where $\Phi_i(z,\bar{z})=\phi_i(z)+\bar{\phi}_i(\bar{z}), i=1,2,3$ are
free fields of $WD_3$, $\lambda$ is the coupling constant of the
perturbation, and $\beta=\sqrt{\frac{p}{p+1}}$. We briefly recall the
method for constructing of nonlocal conserved charges of a perturbed
CFT ([1],[13]). If we assume the existence of some
conserved chiral currents $J(z),\bar{J}(\bar{z})$ for nonperturbed
CFT then for the perturbed currents we have the following
Zamolodchikov's equations up to the first order

\beqa
\bar{\pa}J(z,\bar{z})&=&\lambda \oint \frac{d\om}{2\pi i}\Phi_{pert}
     (\om,\bar{z}) J(z) \label{Zameq} \\
\pa\bar{J}(z,\bar{z})&=&\lambda \oint \frac{d\bar{\om}}{2
\pi i}\Phi_{pert}
     (z,\bar{\om})\bar{J}(\bar{z}) \label{Zameq} \nonumber
\eeqa

If the operator product expansions (OPE) in these contour
integrations have the form

\beqa
\Phi_{pert}(\om,\bar{z}) J(z)&=&\frac{\bar{h}(\om,\bar{z})}{
(\om-z)^{2}}
+ \frac{\pa_{\om}\bar{f}(\om,\bar{z})}{(\om-z)} +regular~terms
\label{OPE} \\
\Phi_{pert}(z,\bar{\om})\bar{J}(\bar{z})&=&\frac{h(z,\bar{\om})}
{(\bar{\om}-\bar{z})^{2}}
+ \frac{\bar{\pa}_{\bar{\om}}f(z,\bar{\om})}{(\bar{\om}-\bar{z})}
+regular~terms \nonumber
\eeqa

then the Zamolodchikov's equations take the form

\beqa
\bar{\pa}J(z,\bar{z})&=&\pa\bar{H}(z,\bar{z}) \label{ZEq} \\
\pa\bar{J}(z,\bar{z})&=&\bar{\pa}H(z,\bar{z}) \nonumber
\eeqa

where

\beqa
\bar{H}(z,\bar{z})&=&\lambda (\bar{h}(z,\bar{z}) +\bar{f}(z,\bar{z}))
\nonumber \\
H(z,\bar{z})&=&\lambda (h(z,\bar{z}) +f(z,\bar{z}))
\nonumber
\eeqa

which means the existence of the conserved charges

\beqa
Q&=&\int\frac{dz}{2\pi i} J + \int\frac{d\bar{z}}{2\pi i} \bar{H}
\label{Cha} \\
\bar{Q}&=&\int\frac{dz}{2\pi i} H + \int\frac{d\bar{z}}{2\pi i} \bar{J}
\nonumber
\eeqa

   For the case under consideration the "maximal" set of currents which
satisfy eqs.\ref{OPE} is

\beqa
J_{1}(z)=\large{e}^{\frac{i}{\beta}(\phi_2(z)-\phi_3(z))};~~~~
\bar{H}_1(z,\bar{z})&=&\lambda\frac{\beta^2}{\beta^2-1}
\large{e}^{i(\frac{1}
{\beta}-\beta)(\phi_2(z)-\phi_3(z))}
\large{e}^{-i\beta(\bar{\phi}_2(\bar{z})
-\bar{\phi}_3(\bar{z}))} \label{Cur} \\
J_{2}(z)=\large{e}^{\frac{i}{\beta}(\phi_1(z)-\phi_2(z))};~~~~
\bar{H}_2(z,\bar{z})&=&\lambda\frac{\beta^2}{\beta^2-1}
\large{e}^{i(\frac{1}
{\beta}-\beta)(\phi_1(z)-\phi_2(z))}
\large{e}^{-i\beta(\bar{\phi}_1(\bar{z})
-\bar{\phi}_2(\bar{z}))} \n
J_{3}(z)=\large{e}^{-i\frac{2}{\beta}\phi_1(z)};~~~~~~~~
\bar{H}_3(z,\bar{z})&=&\lambda\frac{\beta^2}{\beta^2-2}
\large{e}^{-2i(\frac{1}
{\beta}-\frac{\beta}{2})\phi_1(z)}
 \large{e}^{i\beta\bar{\phi}_1(\bar{z})} \n
J_{0}(z)=\large{e}^{\frac{i}{\beta}(\phi_2(z)+\phi_3(z))};~~~~
\bar{H}_0(z,\bar{z})&=&\lambda\frac{\beta^2}{\beta^2-1}
\large{e}^{i(\frac{1}
{\beta}-\beta)(\phi_2(z)+\phi_3(z))}
\large{e}^{-i\beta(\bar{\phi}_2(\bar{z})
+\bar{\phi}_3(\bar{z}))} \nonumber
\eeqa

\newpage

\beqa
\bar{J}_{1}(\bar{z})=\large{e}^{-\frac{i}{\beta}(\bar{\phi}_2(\bar{z})
-\bar{\phi}_3(\bar{z}))};~~~~
H_1(z,\bar{z})&=&\lambda\frac{\beta^2}{\beta^2-1}
\large{e}^{-i(\frac{1}
{\beta}-\beta)(\bar{\phi}_2(\bar{z})-\bar{\phi}_3(\bar{z}))}
\large{e}^{i\beta(\phi_2(z)
-\phi_3(z))} \label{Curb} \\
\bar{J}_{2}(\bar{z})=\large{e}^{-\frac{i}{\beta}(\bar{\phi}_1(\bar{z})-
\bar{\phi}_2(\bar{z}))};~~~~
H_2(z,\bar{z})&=&\lambda\frac{\beta^2}{\beta^2-1}
\large{e}^{-i(\frac{1}
{\beta}-\beta)(\bar{\phi}_1(\bar{z})-\bar{\phi}_2(\bar{z}))}
\large{e}^{i\beta(\phi_1(z)
-\phi_2(z))} \n
\bar{J}_{3}(\bar{z})=\large{e}^{i\frac{2}{\beta}\bar{\phi}_1
(\bar{z})};~~~~~~~~
H_3(z,\bar{z})&=&\lambda\frac{\beta^2}{\beta^2-2}
\large{e}^{2i(\frac{1}
{\beta}-\frac{\beta}{2})\bar{\phi}_1(\bar{z})}
 \large{e}^{-i\beta\phi_1(z)} \n
\bar{J}_{0}(\bar{z})=\large{e}^{-\frac{i}{\beta}(\bar{\phi}_2(\bar{z})+
\bar{\phi}_3(\bar{z}))};~~~~
H_0(z,\bar{z})&=&\lambda\frac{\beta^2}{\beta^2-1}
\large{e}^{-i(\frac{1}
{\beta}-\beta)(\bar{\phi}_2(\bar{z})+\bar{\phi}_3(\bar{z}))}
\large{e}^{i\beta(\phi_2(z)
+\phi_3(z))} \nonumber
\eeqa

with the spins of the charges (\ref{Cha})

\beqa
s_i&=&-\bar{s}_i=\frac{1}{\beta^2}-1;~~~~~~i=0,1,2  \label{Spi} \\
s_3&=&-\bar{s}_3=\frac{2}{\beta^2}-1;   \nonumber
\eeqa

We should recall now that the quasi-chiral components
$\phi_i,\bar{\phi}_i$
of the Toda free fields $\Phi_i$ commute only in the absence of
perturbation $\lambda=0$, but in general their vertex operators
obey certain braiding relations (see ref [13]). Skipping details
we will give the result which easily can be checked by use  of
these braiding relations:the definition of topological charges

\beqa
T_1&=&\frac{\beta}{2\pi}\int_{-\infty}^{\infty} dx\pa_x(\Phi_2-\Phi_3)
     \label{tch} \\
T_2&=&\frac{\beta}{2\pi}\int_{-\infty}^{\infty} dx\pa_x(\Phi_1-\Phi_2)
\n
T_3&=&-2\frac{\beta}{2\pi}\int_{-\infty}^{\infty} dx\pa_x\Phi_1 \n
T_0&=&-2T_2-T_1-T_3 \nonumber
\eeqa

leads to the following algebra of charges:

\beqa
\left[ T_{i},Q_{j}\right] &=&A_{ij}Q_{j} \label{alg} \\
\left[ T_{i},\bar{Q}_{j}\right] &=&-A_{ij}\bar{Q}_{j} \n
Q_{i}\bar{Q}_{j}- q^{-A_{ij}}\bar{Q}_{j} Q_{i}&=&\delta_{ij} a_{i}
(1-q^{2T_{i}}) \nonumber
\eeqa

 where

\beq
A_{ij}=\left(\begin{array}{rrrr}
               2 &0 &-1 &0 \\
               0 &2 &-1 &0 \\
               -1&-1&2  &-2\\
               0 &0 &-2 &4  \end{array} \right) \label{Car}
\eeq

 differs just by diagonal normalization from Cartan matrix of the
affine Lie algebra $A_5^{(2)}$,

\beq
q=e^{-\frac{i\pi}{\beta^2}}
\eeq

is the deformation parameter of quantum group, and

\beqa
a_i&=&\frac{\lambda}{2\pi i}\left(\frac{\beta^2}{\beta^2-1}\right)^2;
{}~~~~~~i=0,1,2 \label{cona} \\
a_3&=&\frac{\lambda}{2\pi i}\left(\frac{\beta^2}{\beta^2-2}\right)^2
\n
\eeqa

The standard substitution ([13],[14])

\beqa
Q_{i}&=&c_{i} e^{s_{i}\th} E_{i} q^{\frac{H_{i}}{2}} \label{tra} \\
\bar{Q}_{i}&=&c_{i} e^{\bar{s}_{i}\th} F_{i} q^{\frac{H_{i}}{2}} \n
H_{i}&=&T_{i} \n
c_{i}^{2}&=&a_{i}(q_{i}^{-2}-1) \n
q_{i}&=& q^{A_{ii}/2} \nonumber
\eeqa

 which introduces the
spectral parameter ($\th$) dependence, transforms the algebra
(\ref{alg}) into the quantum
 affine Lie algebra $A^{(2)}_{5q}$ with Chevalley basis
$E_{i},F_{i},H_{i}$ ($i=0,1,2,3$).

\newpage

\beqa
\left[ H_{i},H_{j}\right]&=&0 \label{algA} \\
\left[ H_{i},E_{j}\right] &=&A_{ij}E_{j} \n
\left[ H_{i},F_{j}\right] &=&-A_{ij}F_{j} \n
\left[ E_{i},F_{j}\right] &=&\delta_{ij}\frac{q^{H_{i}}-q^{-H_{i}}}
{q_{i}-q_{i}^{-1}} \nonumber
\eeqa

The fundamental representation of this algebra can be chosen the same
as for the $A_{5}^{(2)}$ and the basis for Cartan subalgebra in this
representation can be taken as $H_1=diag(1,0,0,0,0,-1)$;$H_2=
diag(0,1,0,0,-1,0)$;$H_3=diag(0,0,1,-1,0,0)$;$H_0=-2H_2-H_1-H_3$.

    The next natural step is the construction of the sixtet
of fundamental soliton fields for the model which will form the
representation of the above written algebra of nonlocal currents.
One of possible candidates can be chosen as $\psi_{i\pm}=
e^{\pm\frac{i}{\beta}\phi_{i}}$, where $i=1,2,3$,
 or another choice :
$\bar{\psi}_{i\pm}=e^{\pm\frac{i}{\beta}\bar{\phi}_i}$.
 It is easily
can be checked by standard technique of conformal field theory that
any of these two sets of fields possesses the correct
 topological charges.

\beqa
 \left[ T_{0},\psi_{1\pm}\right]&=&0,~~\left[
T_{0},\psi_{2\pm}\right]=\pm \psi_{2\pm},~~\left[
T_{0},\psi_{3\pm}\right]=\pm \psi_{3\pm} \label{tzsol}\\
 \left[ T_{1},\psi_{1\pm}\right]&=&0,~~\left[
T_{1},\psi_{2\pm}\right]=\pm \psi_{2\pm},~~\left[
T_{1},\psi_{3\pm}\right]=\mp \psi_{3\pm} \n
 \left[ T_{2},\psi_{1\pm}\right]&=&\pm \psi_{1\pm},~~\left[
T_{2},\psi_{2\pm}\right]=\mp \psi_{2\pm},~~\left[
T_{2},\psi_{3\pm}\right]=0 \n
 \left[ T_{3},\psi_{1\pm}\right]&=&\mp 2\psi_{1\pm},~~\left[
T_{3},\psi_{2\pm}\right]=0,~~\left[
T_{3},\psi_{3\pm}\right]=0 \nonumber
\eeqa

 Clearly each of the set of fields $\psi,\bar{\psi}$ suffers from the
ill-defined action of the part of the charges $Q_{i},\bar{Q}_{i}$
because of the  branch cuts under the contour integrals in part
of the operator product expansions $Q \psi$, and hence does not form
the correct representation of the full algebra (\ref{alg}).
 But actually
we need for our purposes here just the fields which will permit us
to define braiding relations between currents and fundamental
soliton fields which is compatible with the comultiplication
 structure of the
revealed group $A_{5}^{(2)}$.
Using relations (\ref{tzsol}) one can show by the technique of the
braiding relations of the vertices for fields $\phi$ [13],
that for the fields
 $\psi $ defined above, the following braiding relations are valid

\beqa
J_{i}(x)\psi_{j\pm}(y)&=&q^{\tau_{ij\pm}}\psi_{j\pm}(y)J_{i}(x)
\label{bra} \\
\bar{J}_{i}(x)\psi_{j\pm}(y)&=&q^{-\tau_{ij\pm}}\psi_{j\pm}(y)
\bar{J}_{i}(x) \nonumber
\eeqa

for $x<y$ , and commutation of $J_{i}(x)$ and $\psi_{j\pm}(y)$
for $x>y$, where $\tau_{ij\pm}$ are the
topological charges of fields $\psi_{j\pm}$ with respect to $T_{i}$,
and the latter can be read from (\ref{tzsol}).
Such braiding relations induce comultiplication

\beqa
\Delta(Q_{i})&=&Q_{i}\otimes 1 + q^{H_{i}}\otimes Q_{i} \label{copr}\\
\Delta(\bar{Q}_{i})&=&\bar{Q}_{i}\otimes 1 + q^{H_{i}}\otimes
\bar{Q}_{i} \n
\Delta(H_{i})&=&H_{i}\otimes 1 + 1\otimes H_{i} \nonumber
\eeqa

which acts on the tensor product of two soliton states.

 The natural further step in the direction to the construction of
FST of the model is to try to find such an S-matrix for the
fundamental particles (as an operator which acts in the tensor product
 of two vector spaces of representations)
which possesses the symmetry revealed above and expressed by the
algebra of nonlocal charges. That means the S-matrix should
commute with the comultiplication (\ref{copr})

\beq
\left[ S,\Delta(H_{i})\right]=\left[ S,\Delta(Q_{i})\right]=
\left[ S,\Delta(\bar{Q}_{i})\right]=0 \nonumber
\eeq

 Introducing the notation $\hat S=P S$, where $P$ is the permutation
matrix, these equations can be rewritten as

\beqa
\left[ \hat{S}(\theta_{1},\theta_{2}),\Delta(H_{i})\right]&=&0
 \label{coms}\\
\hat{S}(\theta_{1},\theta_{2})
\left( e_{i}\otimes q^{-\frac{H_{i}}{2}}+q^{\frac{H_{i}}{2}}
\otimes e_{i}\right)&=&\left( q^{-\frac{H_{i}}{2}}\otimes e_{i}+
e_{i}\otimes q^{\frac{H_{i}}{2}}\right)\hat{S}(\theta_{1},\theta_{2})
\n
\hat{S}(\theta_{1},\theta_{2})
\left( f_{i}\otimes q^{-\frac{H_{i}}{2}}+q^{\frac{H_{i}}{2}}
\otimes f_{i}\right)&=&\left( q^{-\frac{H_{i}}{2}}\otimes f_{i}+
f_{i}\otimes q^{\frac{H_{i}}{2}}\right)\hat{S}(\theta_{1},\theta_{2})
\nonumber
\eeqa

where $\theta_{1,2}$- rapidities of the incoming particles,

\beq
e_{i}=x_{i}E_{i},~~~f_{i}=x_{i}^{-1}F_i,~~~x_{i}=x_{i}(\theta_{j})=
e^{s_{i}\theta_{j}} \label{spin}
\eeq

and the dependence of $e_{i}$ and $f_{i}$ on the rapidity ($\theta_{1}$
 or $\theta_{2}$) is defined by their positions in tensor product
 (they depend on $\theta_{1}$ on the first place,
 and on $\theta_{2}$ on the
second).

    Such system of equations like (\ref{coms}) was solved with
respect to $S$ (without unitarity and crossing symmetry
conditions) by Jimbo [12] almost for all affine algebras
of symmetry
 of R-matrix $S$, but 
 he used another Cartan basis and his results were obtained in other
gradation. His Cartan basis $h_i$ is connected to our basis $H_i$ by
transformation

\beq
 h_1=H_1-H_2,~~h_2=H_2-H_3,~~h_3=H_2+H_3,~~h_0=-2H_1 \label{carji}
\eeq

with corresponding change in Chevalley generators $E'_i$,$F'_i$,
 such that eq.
 (\ref{coms}) and (\ref{spin}) remain valid in this new basis.
R-matrix was obtained in so called homogeneous gradation, when the
spectral parameter dependence $x$ is introduced as a multiplier of
only one of Chevalley generators, which correspond to the highest
wieght of $A_5$ in its odd component of Dynkin diagramm  automorphism
decomposition - the root number 3 in our case. So in this homogeneous
 gradation

\beqa
e_{3}^{hom}&=&xE'_{3},~~~f_{3}^{hom}=x^{-1}F'_3,~~~
x=x(\theta_{2}-\theta_{1}) \label{hom}\\
e_{i}^{hom}&=&E'_{i},~~~f_{i}^{hom}=F'_i,~~~i=0,1,2 \nonumber
\eeqa

and R-matrix $S$ is a function of $x$.  If we want to make use of
 Jimbo's result for the $A_{5}^{(2)}$  R-matrix in our S-matrix
construction, we should change homogeneous gradation into spin,
(which, as we saw, was naturally dictated by the nonlocal charges of
the system) by "gauge" transformation of the Jimbo's solution:

\beq
\tilde{R}(x,k)=\s_{21} R(x,k) \s_{12}^{-1} \label{gaug}
\eeq

where

\beq
\s_{12}=x_{0}(\th_{1})^{-\sum_{i=1}^{3}\frac{h_{i}a_{i}}{2}}\otimes
x_{0}(\th_{2})^{-\sum_{i=1}^{3}\frac{h_{i}a_{i}}{2}} \label{sig}
\eeq

Using  obvious relations
\newpage

\beqa
y^{h_{i}/2} E'_{j} y^{-h_{i}/2}&=& y^{a_{ij}/2} E'_j \label{rche}\\
y^{h_{i}/2} F'_{j} y^{-h_{i}/2}&=& y^{-a_{ij}/2} F'_j \nonumber
\eeqa

we have the following system of equations which fixes $a_i$ in
(\ref{sig}):

\beqa
&x_{0}&^{a_{1}-\frac{a_{2}}{2}-\frac{a_{3}}{2}}=x_0 \label{sysa}\\
&x_{0}&^{-\frac{a_{1}}{2}+a_{2}}=x_{0} \n
&x_{0}&^{-\frac{a_{1}}{2}+a_{3}}=x_0 \n
&x_{0}&^{-a_{1}}x=x_1 \nonumber
\eeqa

 Solving first 3 eqations we have $a_{1}=4$, $a_{2}=a_{3}=3$,
and the last equation gives us important relation

\beq
x=x_{1} x_{0}^{4} \label{xx1}
\eeq

Since the dependence on the coupling constant $\beta$ enters
Jimbo's R-matrix only through its dependence on $x$, the relation
 (\ref{xx1}) gives us the effective coupling constant $\xi$ as
function
of $\beta$:

\beq
x=e^{\frac{2\pi\theta}{\xi}},~~~~\frac{2\pi}{\xi}=\frac{6}{\beta^2}
-5 \label{effcon}
\eeq

We could now take "gauged" R-matrix solution in the spin
 gradation and,
multiplying it by some scalar function of $x$ and $k$, imply
 unitarity
and crossing symmetry conditions, trying to construct the S-matrix.
Let us make few comments on this way of S-matrix construction.
We checked that Jimbo's R-matrix solution, being gauged
from the homogenious
to the spin gradation with constants $a_i$ found above,
 becomes crossing
invariant, since the multiplying factor in the crossing
 relation exactly
cancels by the "gauge" factor.

  If we wuold like now to fit the deformation parameter
 $k$ with the coupling
constant by use of the crossing transformation for the
 R-matrix solution
($x\rightarrow -\frac{x}{k^6}$ corresponds
to $\th\rightarrow i\pi -\th$),
then in the notations

\beq
x=e^{-i\pi a},~~~~a=\frac{2i\th}{\xi},~~~~k=e^{-i\pi b} \nonumber
\eeq

we get

\beq
b=-\frac{\pi}{3\xi}+\frac{2m+1}{6} \nonumber
\eeq

where $m$ is an arbitrary integer number, and  choice of it
in principle
is crutial for the S-matrix pole structure. In addition
there is well known
CDD ambiguity in the solution of unitarity and crossing
 symmetry conditions
for the scalar factor mentioned above. Both of these two
 ambiguities (CDD one
 and the integer $m$ choice) was reduced in the work
\cite{3} in the case of
$A_2^{(2)}$ S-matrix construction for $(1,2)$-perturbed
Virasoro minimal models
 by comparison of the S-matrix for lightest breather-breather
 scattering, found
by bootstrap from the kink-kink Smatrix, with the known
 solution for the
real coupled Toda S-matrix \cite{15}. Such a pattern for
 comparision exists in
our case as well -- there is the S-matrix solution for the
 real coupled
$A_{2n-1}^{(2)}$ Toda model \cite{10}, and we can try to do the same,
but we will  discuss this problem elsewhere.


    \section{Discussion}

   We have shown by counting argument
that $(21...1|11...1)$-perturbations of $WD_{n}$ theories
(and  $(21|11)$ - of $W_3$) are integrable, and it was not
surprizing after we realized  their $B_n$  (and $G_2$)
imaginary coupled ATFT structure on the "classical level".
The conjectured $A_{2n-1}^{(2)}$ symmetry of their  quantum
S-matrix has been proved by explicit construction of  nonlocal charges
with this quantum group symmetry    and it was shown that there are
a set of fields, which play the role of fundamental solitons
in the sense
that their braiding relations with nonlocal charges give rise to the
correct comultiplication for this quantum group.

We saw that the R-matrix of the model, as a commutant of the found
coproduct structure, should be connected   to the known $A_{2n-1}^{(2)}$
 R-matrix Jimbo's solution   by change of gradation. As we
pointed out,in  the spin gradation, dictated by the revealed current
structure, the Jimbo's R-matrix becomes crossing invariant, which gives
a hope to expect that it can serve as a  good basis for the S-matrix
construction. In addition, the consistency condition of the gradation
change fixed explicitly the S-matrix effective coupling constant
dependence.

It should be emphasised here, that the considered model was not
a minimal W model, since we started from the free bosonic representation
with central charge $c=n$ adding to it some screening operators
and perturbation. The presense of the "curvature" term of
Feigin-Fucks minimal model construction will change all the picture
drasticaly, but as we know from the  examples of Sine-Gordon and ZMS
models, the perturbation of considered minimal W-models probably
 can be obtained as
a quantum group reduction of the S-matrix discussed here.
However it seems that the most natural and the simplest way
of the S-matrix construction for the  perturbed minimal models
can be done (\cite{18}) on the base of the RSOS models, which are partialy
studied for our groups of symmetry (\cite{16},\cite{17}).

The examples of integrable perturbations of W-invariant theories
analyzed in this paper probably don't exhaust all of them, which
reveal more rich structure then in Virasoro case. It can be seen
by comparizon of Dynkin
diagrams of affine Lie algebras with those of non affine Lie algebras
of other type $X$. Each case, when
 the  former one can be obtained from the latter $X$
 by adding to $X$ some combination of
its fundamental weights, might be considered as a candidate for
integrability of the perturbation of $WX$ by the field corresponding
to this specific combination of wheights of $X$.
As examples we can mention here $(3,1,...1)$ perturbation of $WD_n$
theories, which gives the $A_{2n-1}^{(2)}$ Affine Toda theory
(this perturbation is irrelevant for the unitary minimal models,
but is relevant in some nonunitary minimal models),and, probably
more attractive, example of vector perturbation of $WB_n$ theories,
which seems to give the $A(0,2n)^{(4)}$ supersymmetric ATFT
(with broken supersymmetry).
 But each of these cases requires
separate detailed investigation.

 \section{Acknowledgements}

I  thank S.Elitzur and I.Vaysburd for helpful
discussions and C.Efthimiou  for useful comments on his work
 [14].

\end{document}